\documentclass{emulateapj}
\usepackage{amssymb,amsmath}
\usepackage{graphicx}
\usepackage{natbib}
\begin{document}
\title{Spins of large asteroids: Hint on a primordial distribution of their spin rates}
\author{Elad Steinberg and Re'em Sari}

\begin{abstract}
The Asteroid Belt and the Kuiper Belt are relics from the formation of our solar system. Understanding the size and spin distribution of the two belts is crucial for a deeper understanding of the formation of our solar system and the dynamical process that govern it.
In this paper, we investigate the effect of collisions on the evolution of the spin distribution of asteroids and KBO's.

We find that the power law nature of the impactors' size distribution leads to a L\'evy distribution of the spin rates. This results in a power law tail of the spin distribution, in stark contrast to the usually quoted Maxwellian distribution.
We show that for bodies larger than 10 km, collisions alone lead to spin rates peaking at 0.15-0.5 revolutions per day.
Comparing that to the observed spin rates of large asteroids ($R>50$ km), we find that the spins of large asteroids, peaking at $\sim1-2$ revolutions per day, are dominated by a primordial component that reflects the formation mechanism of the asteroids.
Similarly, the Kuiper Belt has undergone virtually no collisional spin evolution, assuming current day density. Collisions contribute a spin rate of $\sim0.01$ revolutions per day, thus the observed fast spin rates of KBOs are also primordial in nature.
\end{abstract}
\section{Introduction}
It has long been suggested that the observed properties of the Main Belt and the Kuiper Belt give us a unique glimpse into the infant stages of our solar system and valuable information about the formation of terrestrial planets \citep{TZ80,DM82,B2005,SS07}. The current observed spin distribution in the Main-Belt has been claimed to fit a Maxwellian distribution, or a combination of such, \citep{HB79,3M,Pravec2000} due to the random accretion onto the asteroids during their formation. Although a Maxwellian fits the spin distribution reasonably well, it predicts too few slow and fast rotators compared to the observed population \citep{Pravec2000,DP2009}. If the spins of bodies are gained by high velocity impacts (as is the case today in the solar system), then the distribution should not be Maxwellian since collisions have much more specific energy than the maximum rotation rate for gravitationally bound bodies.

Theoretical work on planet formation predicts that protoplanets formed from a uniform cold disk of planetesimals, end up with a slow retrograde spin \citep{LK91,DT93,L97}. After the disk has been depleted the giant impact phase commences, where spin is acquired by random large impacts \citep{A99,C01,GLS04}. The spin gained due to giant impacts does not give rise to a Maxwellian distribution since the angular momentum is dominated by few large impacts. However, protoplanets might also accrete via semi-collisional accretion \citep{SS07,J2010}, where protoplanets form an optically thin accretion disk that is accreted onto the protoplanet. The accretion from a disk gives rise to fast prograde spin rates. Determining the amount of angular momentum that the largest asteroids have gained during the lifetime of the solar system from collisions, can help pinpoint the primordial spins of asteroids and therefore differentiate between the mechanisms of gaining angular momentum,i.e. semi-collisional accretion, giant impacts or angular momentum from the initial cold disk.

\section{Pure Collisional Evolution}
In the following section we assume that the impact velocities are always much greater than the target's escape velocity; hence, all collisions are erosive rather than accreting collisions.
The calculated distribution arises from collisions only, assuming zero initial spin. If an initial spin distribution is present, then the spin distribution after collisions will simply be a convolution of the initial spin distribution and our calculated result.
\subsection{Ideal Collisions}
The simplest (naive) approach to model the change in spin from a single collision gives
\begin{equation}
\Delta\vec{w}=m_p\frac{\vec{R}_t\times\vec{v}}{I}
\label{eq:ideal}
\end{equation}
where $\Delta\vec{w}$ is the change in the spin vector, $m_p$ is the mass of the projectile, $\vec{R}_t$ is the location of the impact relative to the asteroid's center of mass and whose magnitude is the radius of the target, $\vec{v}$ is the impact velocity and $I$ is the target's moment of inertia.
The following derivation is similar in its structure to the calculation preformed in \cite{RB}.

The rate of collisions that bestow upon the target a change in spin with a magnitude between $w'$ and $w'+dw'$, $\mathcal{R}(w')dw'$ is given by
\begin{equation}
\mathcal{R}(w')=P_i\int \delta(w'-|
\vec{\Delta w}|)f(v)n(R_p)R_t^2\sin(2\theta)dvdR_pd\theta
\label{eq:integral}\end{equation}
where $f(v)=\sqrt{\frac{2}{\pi}}\frac{v^2\exp{(-v^2/2v_0^2)}}{v_0^3}$ is the PDF of the impactors velocities, $n(r)=\frac{N_0(\alpha-1)}{R_0}\left(\frac{R_0}{r}\right)^\alpha$ is the differential size distribution of the impactors and $\theta$ is the angle between the impactor's velocity and the radius vector of the target. $P_i$, called the intrinsic collision probability \citep{B1994}, is defined such that the probability per unit time that a target will be hit is given by $P_iR_t^2N_{imp}$. A simple way to estimate this number is
\begin{equation}
P_i\approx\left(T({a_{max}^2-a_{min}^2})\right)^{-1}
\end{equation}
where $T$ is the orbital period and $a_{max}$ and $a_{min}$ are the edges of the asteroid belt.
With $a_{min}=2.06\;\textrm{AU}$, $a_{max}=3.27\;\textrm{AU}$ and evaluating $T$ at the center of the belt, our estimate yields $P_i=3.6\times10^{-18}\textrm{km}^{-2}\textrm{year}^{-1}$.
\cite{B1994} estimated the value of $P_i$ to be $P_i=2.86\times10^{-18}\textrm{km}^{-2}\textrm{year}^{-1}$ by numerically calculating intersections between known orbits of asteroids. This result is in good agreement with our less precise calculation.

For simplicity, we assume that the velocity distribution is isotropic. In reality,
the velocity dispersion within the plane is about a factor of 1.5 lower than
the velocity dispersion perpendicular to the plane. This induces some
anisotropy in the spin axis distribution. In principle, one can conduct a similar
analysis to the one presented here for the spin distribution in each axis separately.
However, since observation of the spin axis are more sparse, and the velocity
anisotropy is less than a factor of two, we leave this complication for future research.

Integrating eq. \eqref{eq:integral} yields
\begin{equation}
\begin{split}
\mathcal{R}(w')=&N_0P_iR_t^2\frac{R_0}{v_0}\frac{(\alpha-1)}{\sqrt{\pi}(5+\alpha)}\Gamma\left(\frac{8+\alpha}{6}\right)
\frac{5^\frac{\alpha-1}{3}}{2^\frac{\alpha-13}{6}}\times\\
&\left(\frac{R_0}{R_t}\right)^\frac{4(\alpha-1)}{3}
\left(\frac{v_0}{R_0w'}\right)^\frac{\alpha+2}{3}.
\end{split}
\end{equation}
The frequency per unit of spin space ($d^3w$) at which an object of size $R_t$ with spin rate $\vec{w}$ is perturbed into a spin rate of $\vec{w}+\vec{w}'$ is defined to be $p(w')$
which is related to $\mathcal{R}(w)$ via
\begin{equation}
\mathcal{R}(w')=4\pi w'^2p(w').
\end{equation}
The evolution of the spin distribution is given by the collisional Boltzmann equation
\begin{equation}
\frac{\partial f(w,t)}{\partial t}=\int p(w')[f(|\vec{w}-\vec{w'}|,t)-f(w,t)]d^3w'
\label{eq:boltzmann}
\end{equation}
where $f$ is the PDF of the spin vector normalized such that $\int f(w',t)d^3w'=1$.

This equation can be solved by a self-similar solution of the type
\begin{equation}
f(w,t)=F(t)g(w/w_c(t)).
\label{eq:anstaz}
\end{equation}
The normalization criteria dictates that $F(t)=w_c^{-3}(t)$ and $\int g(x)d^3x=1$ where $x=w/w_c(t)$. With our self-similar ansatz eq.\eqref{eq:boltzmann} becomes
\begin{equation}
3g(x)+x\frac{dg}{dx}+\frac{1}{\pi^2}\int x'^{-(8+\alpha)/3}\left[g(|\vec{x}-\vec{x}'|)
-g(x)\right]d^3x'=0
\label{eq:integ}
\end{equation}
and
\begin{equation}
\begin{split}
\frac{\dot{w_c}(t)}{w_c(t)}=&N_0P_iR_t^2\sqrt{\pi}\frac{(\alpha-1)}{5+\alpha}\Gamma\left(\frac{8+\alpha}{6}\right)
\frac{5^\frac{\alpha-1}{3}}{2^\frac{\alpha-1}{6}}\left(\frac{R_0}{R_t}\right)^
\frac{4(\alpha-1)}{3}\times\\
&\left(\frac{v_0}{R_0w_c(t)}\right)^\frac{\alpha-1}{3}.
\end{split}
\end{equation}
The typical spin rate can be easily found to be
\begin{equation}
\begin{split}
w_c(t)=&\left(N_0P_iR_t^2\sqrt{\pi}\frac{(\alpha-1)^2}{3(5+\alpha)}\Gamma\left(\frac{8+\alpha}{6}\right)
\frac{5^\frac{\alpha-1}{3}}{2^\frac{\alpha-1}{6}}\left(\frac{R_0}{R_t}\right)^
\frac{4(\alpha-1)}{3}\times\right.\\
&\left.\left(\frac{v_0}{R_0}\right)^\frac{\alpha-1}{3}t\right)^\frac{3}{\alpha-1}.
\end{split}
\end{equation}
This result is roughly equal to the spin imparted from a single collision by the largest impactor that has a probability of order unity to hit the target within a time $t$.
Since eq.\eqref{eq:integ} is linear in $g$, a Fourier transform method for the solution is viable. Taking the Fourier transform defined as $G(k)=\int g(x)e^{i\vec{k}\cdot\vec{x}}d^3x$ transforms eq.\eqref{eq:integ} to
\begin{equation}
\frac{dG(k)}{dk}=\frac{4}{\pi}\sin\left(\pi\frac{\alpha+8}{6}\right)\Gamma\left(
-\frac{2+\alpha}{3}\right)k^\frac{\alpha-4}{3}G(k),
\end{equation}
whose solution is, taking into account the normalization $G(0)=1$, is
\begin{equation}
G(k)=e^{-\beta k^\gamma}
\end{equation}
where
\begin{eqnarray}
\beta&=&\frac{12}{\pi(\alpha-1)}\sin\left(\pi\frac{\alpha+8}{6}\right)\Gamma\left(
-\frac{2+\alpha}{3}\right)\nonumber\\
\gamma&=&\frac{\alpha-1}{3}.
\end{eqnarray}
The inverse transform can only be done analytically for the special case of $\alpha=4$ where we have
\begin{equation}
g(x)=\frac{1}{\pi^2(1+x^2)^2}
\label{eq:cauchy}
\end{equation}
which is the three dimensional Cauchy distribution.
For other slopes of the size distribution, the inverse transformation must be calculated numerically. However, the tail of the distribution, where $x\gg1$ always has the form of $g(x)=3x^{-\frac{8+\alpha}{3}}/(\alpha-1)\pi^2$. This power law tail arises from the fact that fast spins are a consequence of rare events where the target had only one extremely rare large impact. The distribution is always a L\'evy distribution, which has a long tail, unlike the Maxwellian distribution which has an exponential cutoff.
\begin{figure}
    \plotone{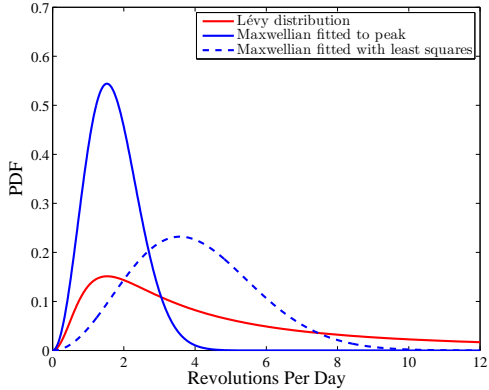}
  \caption{Comparison between the L\'evy distribution plotted with a red solid line and Maxwellian distributions plotted with blue lines. The L\'evy distribution was plotted using $\alpha=3.5$ with $t=4.5\cdot10^9$ years and $N_0=10^6$.}\label{fig:maxwell_fit}
\end{figure}
In fig. \ref{fig:maxwell_fit} we overlay the L\'evy distribution and two Maxwellian distributions, one fitted to have the same location of the peak and the other fitted with a least squares method.

There is a clear difference between the two Maxwellian distributions and the L\'evy distribution. The presence of the long tail is very distinct. The general solution of eq. \eqref{eq:anstaz} has similar properties the solution of the special case where $\alpha=4$ given in eq. \eqref{eq:cauchy}. The long tail is dominated by single, large and rare events that contribute almost all of the angular momentum while a Maxwellian distribution would rise from a series of small and common events.
\subsection{Non Ideal Collisions}
Equation \ref{eq:ideal} neglects the change in the spin vector due to mass ejected, which can change the moment of inertia as well as carry away angular momentum. \cite{AMD} have shown that when the ejected mass, $\Delta M_t$, is small compared to the target's mass, $M_t$, the ejected material tends to carry away with it angular momentum that can be roughly estimated as
\begin{equation}
\Delta\vec{w}\approx -\frac{\Delta M_t}{M_t}\vec{w}
\end{equation}
A simple way to understand this expression is as follows. If an asteroid is rotating with a spin $\vec{w}$ and is hit with a projectile, the ejected material will be ejected in a cone around the impact site. All of the mass that got ejected with velocity greater than $v_{esc}+wR_t$ is lost and all of the ejecta with velocities less than $v_{esc}-wR_t$ fall back onto the target. The net loss of angular momentum arises from the difference in velocity between the two extremes of the ejecta and the fact that most of the mass is ejected at close to the escape velocity.
This angular momentum loss called the Angular Momentum Drain effect, was extended to the case of catastrophic impacts by \cite{splash}. Since the effect is proportional to the target's spin rate, it will be more effective in slowing down fast spinning objects. This effect becomes important when the angular momentum gained by the impact is roughly equal to that lost due to the ejecta.
The amount of ejected mass for hyper-velocity impacts (where the impactor is much smaller than the target) can be roughly estimated as \citep{BE99,SL09}
\begin{equation}
\frac{\Delta M_t}{M_t}\approx \frac{m_pv^2}{4M_tQ_D}
\end{equation}
where $Q_D$ is the energy per unit target mass that is needed to disperse half of the target's mass to infinity.
The spin rate where Angular Momentum Drain starts to dominate is
\begin{equation}
w_{AMD}\approx\frac{4Q_D}{vR_t}\approx 9.2\left(\frac{R_t}{\textrm{km}}\right)^{0.36}\frac{\textrm{revolutions}}{\textrm{day}}
\end{equation}
where we have taken values for $Q_D$ from \cite{BE99} for a basalt target and an impact velocity of $5$ km/s. For ice targets and rubble piles the value of $Q_D$ will be lower \citep{BE99,SL09,LS092}.

A further complication may arise from the fact that for impacts with $\Delta M_t/M_t \gtrsim 0.1$ the efficiency of angular momentum transfer between the projectile and the target is not constant and is difficult to model \citep{LA97,DirectN2000,AMeff07,AMeff09}. For even larger collisions, the target might be completely disrupted and broken into smaller fragments.

Small asteroids in the Main-Belt ($R\le10$ km) might have their spin altered by the YORP effect \citep{YORP}, where asymmetrical reflection of solar radiation/thermal emission can torque an asteroid. Although the timescale of YORP is relatively fast ($\sim10^7$ years for a $R=1$ km asteroid with a timescale scaling as $R^2$), it is not clear that small crater inducing collisions (on the order of $\sim 0.1R$ \citep{SS11}) can not prolong this time scale by randomly changing the YORP sign.

Very large impacts can drastically change the size of an asteroid, which breaks our assumption of a constant size. However, \cite{B05b} have shown that asteroids with radii $R\gtrsim 5$ km have a collisional lifetime which is larger than the age of the solar system. We limit ourselves to calculate only asteroids whose collisional lifetime is larger than the age of the solar system, thus ensuring a roughly constant size.

We neglect all of the above complications and compute the spin distribution for a collisional system bearing in mind that we overestimate the population with fast spins (above $w_{AMD}$ or resulting from impacts with $\Delta M_t/M_t \gtrsim 0.1$).

\section{Main-Belt Observed Spin Rates}
\subsection{Observational Parameters}
We obtain spin rates for Main-Belt asteroids from the Asteroid Lightcurve Database \citep{LCDB}. From this database we took only asteroids that were given a quality flag $2-$ and above, which represents good quality spin periods, and with no known family association. Since it is difficult to get a spin rate measurement for asteroids that revolve slower than once per day, there is a deficit of slow rotators in the reported literature \citep{Pravec2000,Pravec2008}. On the other hand, for fast spinning Near Earth Asteroids, \cite{Pravec2000} have shown that the change in the magnitude due to the spin decreases as the spin rate increases. Note that the observational biases are not significant for large asteroids. For asteroids with $R>25$ km about $96\%$ have known spin rates and for $R>15$ km about $94\%$ have known spin rates.

Our estimates for the asteroid size distribution for asteroids with $H\le15$ is taken from the JPL website and we assume an albedo of $p_V=0.09$ for asteroids with no known albedo. For asteroids in the size range of $0.15\;\textrm{km}\le R\le2.23\;\textrm{km}$ we extend the size distribution with a differential slope index of $\alpha=2.5$ taken from the observation of \cite{G09} and confirmed by \cite{WISE}. Asteroids smaller than $R\le0.15\;\textrm{km}$ are extended with a differential slope index of $\alpha=3.7$ corresponding to the predicted slope of basalt in the strength regime \citep{BE99,OG2003,OG2005,B2005}.

For a given radii bin, we take the mean radius and calculate the largest impactor, $R_0$, that likely occurred during the lifetime of our solar system
\begin{equation}
N(r>R_0)P_iR_t^2t=1
\end{equation}
where $N(r>R_0)$ is the cumulative number of asteroids, $P_i=2.86\cdot10^{-18}\;\textrm{km}^{-2}\;\textrm{year}^{-1}$ taken from \cite{B1994} and $t=4.5\cdot10^9$ years. We then calculate the $N_0$ and $\alpha$ corresponding to $R_0$ from our size distribution. This is done since the real size distribution in the Main Belt is not a perfect power law, but rather has ``waves" imposed on the power law. Our calculated values of $N_0,R_0$ and $\alpha$ are presented in table \ref{table:param}.
\begin{table}
\centering
\begin{tabular}{|c|c|c|c|}
  \hline
  Target Size Bin & $N_0$ & $R_0(\textrm{km})$ & $\alpha$ \\
  \hline
  10-17 km & $4.6\times10^5$ & 0.67 & 2.5 \\
 18-31 km & $1.3\times10^5$ & 1.51 & 2.5 \\
 31-53 km & $4.6\times10^4$ & 2.79 & 3.29 \\
  54-92 km & $1.6\times10^4$ & 4.3 & 3.52 \\
  \hline
  \end{tabular}
  \caption{The calculated values for the prefactor and slope of the size distribution of the impactors for different target sizes.}
  \label{table:param}
\end{table}
Using $v_0=3.32\;\textrm{km}\;\textrm{sec}^{-1}$ (giving a mean impact velocity of $5.3\;\textrm{km}\;\textrm{sec}^{-1}$ \citep{B1994}) allows us to calculate the spin that was gained by collisions and compare it to the observed population.
\subsection{Comparing Observations with Theory}
We compare observations with the expected distribution from collisions. The histograms of spins overlaid with the expected contribution from collisions are plotted for different target sizes in figures 2-5. Also plotted is the expected contribution from collisions but trimmed at 6 revolutions per day and normalized accordingly. The latter best represents the contribution from only ``physical" collisions, where the impact is not large enough to spin an asteroid (with a density of 2 $\textrm{g}/\textrm{cm}^3$) above its maximum allowed spin \citep{chandra}.
\begin{figure}
\plotone{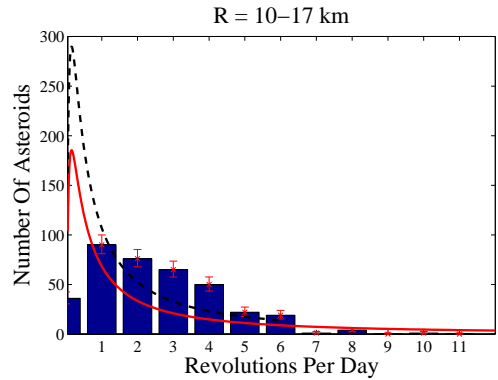}
\caption{Histogram of known spin rates of asteroids with radii in the range $10\;\textrm{km}\le R\le 17\;\textrm{km}$. The red solid line is the solution to eq.\eqref{eq:boltzmann} while the black dashed line is the same as the red line but without fast spinners (6 revolutions per day) and normalized accordingly. Error bars denote $1\sigma$ Poisson noise.}
\label{fig:f1}
\end{figure}

In all of the different size bins, it is clear that there is a big deficit of slow rotators (0-1 revolutions per day) and an excess of intermediate rotators (2-4 revolutions per day) compared to what collisions would have imparted. One plausible explanation is the fact that asteroids are ``born" (either as primordial bodies or during a fragmentation event) with some initial spin. The observed population is thus a convolution between the initial spin and the spin that is imparted by collisions.

\cite{B05b} claim that asteroids with $R\ge55$ km are primordial in the sense that they are not fragments of larger bodies but rather remnants of the accretion process that occurred during the infancy of our solar system. Due to this, we focus our attention of the largest asteroids, which are remnants of the early solar system.

\begin{figure}
\plotone{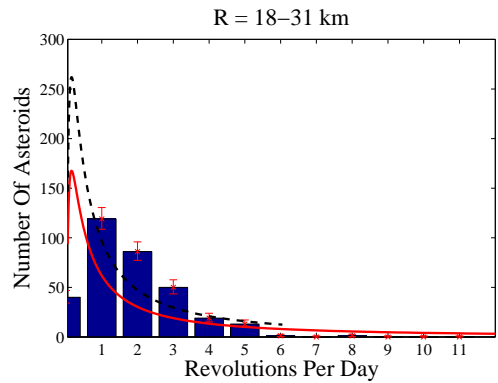}
\caption{Histogram of known spin rates of asteroids with radii in the range $18\;\textrm{km}\le R\le 31\;\textrm{km}$. The red solid line is the solution to eq.\eqref{eq:boltzmann} while the black dashed line is the same as the red line but without fast spinners (6 revolutions per day) and normalized accordingly. Error bars denote $1\sigma$ Poisson noise.}
\label{fig:f2}
\end{figure}

\begin{figure}
\plotone{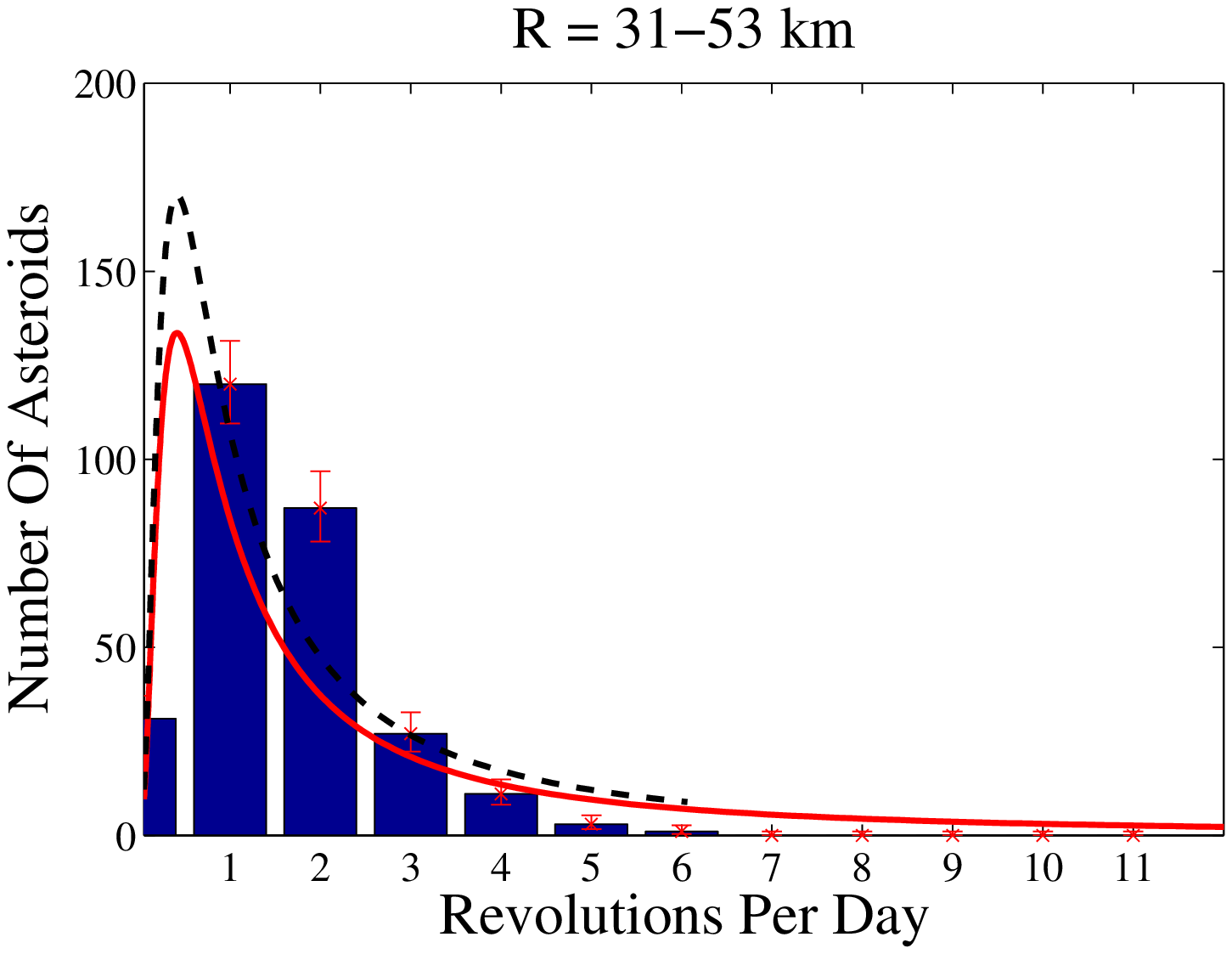}
\caption{Histogram of known spin rates of asteroids with radii in the range $31\;\textrm{km}\le R\le 53\;\textrm{km}$. The red solid line is the solution to eq.\eqref{eq:boltzmann} while the black dashed line is the same as the red line but without fast spinners (6 revolutions per day) and normalized accordingly. Error bars denote $1\sigma$ Poisson noise.}
\label{fig:f3}
\end{figure}
\begin{figure}
\plotone{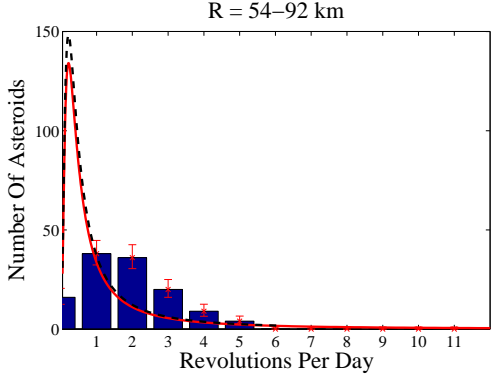}
\caption{Histogram of known spin rates of asteroids with radii in the range $54\;\textrm{km}\le R\le 92\;\textrm{km}$. The red solid line is the solution to eq.\eqref{eq:boltzmann} while the black dashed line is the same as the red line but without fast spinners (6 revolutions per day) and normalized accordingly. Error bars denote $1\sigma$ Poisson noise.}
\label{fig:f4}
\end{figure}

In our the size range of 54-92 km, the deficit (relative to the distribution with a cutoff at 6 revolutions per day) of predicted spins in the range $2-5$ revolutions per day is 44 asteroids.
Since these asteroids are most likely to be primordial, the observed excess of intermediate rotators can only be of primordial origin.

\begin{figure}
\plotone{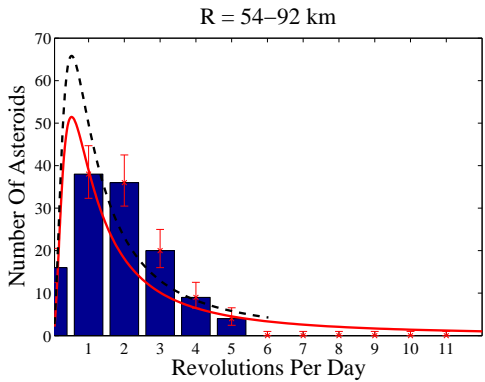}
\caption{Histogram of known spin rates of asteroids with radii in the range $54\;\textrm{km}\le R\le 92\;\textrm{km}$. The red solid line is the solution to eq.\eqref{eq:boltzmann} with a pseudotime of $10^{10}$ years while the black dashed line is the same as the red line but without fast spinners (6 revolutions per day) and normalized accordingly. Error bars denote $1\sigma$ Poisson noise.}
\label{fig:f5}
\end{figure}

\cite{B2005} have shown that if the Asteroid Belt had for a short time a higher density, the effect of the increased density on collisions can be incorporated by integrating the system further in time with a regular density. This ``Pseudotime" is just the addition of the increased density fraction times its duration added to the solar system's age.
In order to support our claim of primordiality, we replot fig. \ref{fig:f4} in fig. \ref{fig:f5}, but instead of using the age of the solar system we use a pseudotime of $t=10^{10}$ years in order to include the possibility of increased collision rate due to the Main-Belt being more populated before Jupiter was fully formed. The observed excess for this scenario is reduced to 17, thus a denser belt lessens the discrepancy between theory and observations but does not fully resolve it. The observed rotation rates of the largest asteroids must have come from primordial origin and could have not come from collisions.

\section{Kupier Belt}
Due to the paucity of observational data on the spins of Kuiper Belt Objects, we only give estimates for the contribution of collisions without comparing histograms.

Since there are several components to the Kuiper Belt (classical, scattered etc.), every region has its own physical properties. In order to simplify the calculation, we assume a uniform Kuiper Belt with the following physical properties.
The size distribution is assumed to be a power law with $N_0=1.8\cdot10^9$, $R_0=1$ km and $\alpha=3.77$, this is calculated by takeing the observed size distribution at sizes 250 meters and 10 km from \cite{Hilke2012,Hilke2013} and interpolating between them while assuming that the Kuiper Belt has a maximum latitude of $10^\circ$. The mean collision velocity is taken to be 0.5 km/s and  the intrinsic collision probability to be $P_i=1.3\cdot10^{-21}\textrm{km}^{-2}\textrm{year}^{-1}$ taken from \cite{KBOPI2}.

\begin{figure}
\plotone{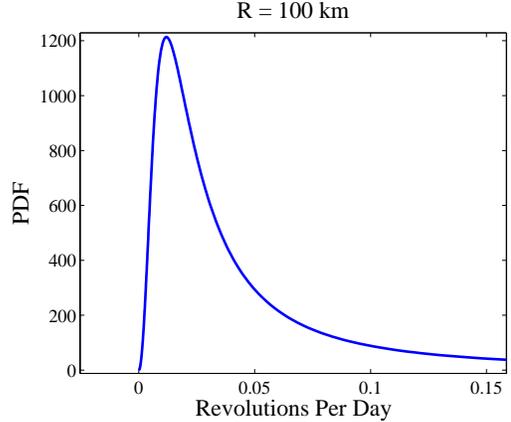}
\caption{The PDF of the calculated spin rates from collisions in the Kuiper Belt for targets with $R=100$ km. The peak of the distribution is well below the typical spin rates of known KBO's.}
\label{fig:f6}
\end{figure}

In fig. \ref{fig:f6} we plot the calculated spin distribution for 100 km objects. The typical spin rate is very low only $\sim10^{-2}$ revolutions per day compared to a few revolutions per day in the observed population. The Kuiper Belt has virtually undergone no collisional spin evolution and the observed spins are primordial.

If we have used more recent estimates of the Kupier Belt collision probabilities \citep[e.g.]{KBOPI}, the typical spin rates would have been even lower.

Future observations of the Kuiper Belt that will extend our known spin population will shed much light on the formation processes of the early solar system due to the lack of collisional spin evolution of the Kuiper Belt.
\section{Discussion}
Asteroids with radii $10-53$ km have a clear deficit of fast rotators compared to what should have been obtained from collisions. This suggests that the efficiency of angular momentum transfer in energetic impacts has to be less than unity. The deficit can not be a result of the Angular Momentum Drain effect, since the fast spin was bestowed by a large, rare impact event. Subsequent large, rare impact events that are needed for the Angular Momentum Drain effect, are unlikely.

The observed excess of intermediate spinners and deficit of slow spinners for the smaller asteroids can be explained by having asteroids ``born" (either primordial or from a fragmentation process) with an initial spin rate that is of order a revolution per day.

The maximal spin rate (retrograde) that arises from accreting the angular momentum of planetesimals in a Keplerian disk with no velocity dispersion has been shown
by \cite{DT93} to be
\begin{equation}
f\approx 1\sqrt{\frac{\textrm{2.5\;AU}}{a}}\left(\frac{\rho}{\textrm{2 g/cm}^3}\right)^{1/3}\;\frac{\textrm{revolutions}}{\textrm{day}}
\end{equation}
where $a$ is the semi-major axis of the object.
Assuming a constant impactor size, the case of stochastic accretion gives rise to an object that spins close to the breakup spin for a single accretion event and decreases as the square root of the number of events that were involved in the stochastic accretion. However, stochastic accretion should also give rise to a uniform distribution in the cosine of the obliquity which is not observed \citep{H2013}.
In the asteroid belt, it is apparent that there are too many fast spinning asteroids to be explained by both ordered and stochastic accretion as well as collisions during the lifetime of the solar system.

The faster population of the large asteroids must have acquired their spin by another mechanism, perhaps via semi-collisional accretion and then later evolution by collisions during the lifetime of the solar system. Supporting this claim is the observation that there is a preference for prograde rotation for asteroids with $D>60$ km \citep{H2013}. Semi-collisional accretion is the only proposed evolution path of large asteroids that can account for fast prograde spins. From the data in \cite{H2013} it is apparent that about two thirds of these large asteroids are prograde. Assuming that all of the asteroids that spin too fast compared to what would arise from collisions in fig \ref{fig:f4} originate from semi-collisional accretion and that they are prograde while the rest have no preference; give us a ratio of prograde to retrograde that is consistent with the observations of the large asteroids.

Another possibility is that the Main-Belt has a pseudotime of about $10^{10}$ years (e.g. by having 550 times the current mass for 10 million years) and the large asteroids acquired most of their spin by stochastic accretion. Since large asteroids have a preference for prograde spins \citep{H2013} while collisions and stochastic accretion have no preference support the former scenario.

The largest objects in the Kuiper Belt (with the exception of Pluto which is a slow rotator and Haumea which is a very fast retrograde rotator) typically spin with $1-3$ revolution per day. Determining the collisional evolution of the Kuiper Belt is more challenging since it is unclear when and where it was formed. Our analysis suggests that either the Kuiper belt was once much more dense and/or the spin distribution is a relic of the formation mechanism. Future efforts to observe spins of Kuiper Belt objects with care to minimize the observational bias will shed light on the formation of the Kuiper Belt.

\begin{acknowledgments}
ES is supported by an Ilan Ramon grant from the
Israeli Ministry of Science. This research is supported in part by  ISF, ISA, iCORE grants and a Packard Fellowship.
\end{acknowledgments}
\bibliography{LevyPaper}
\end{document}